\documentclass[sts]{imsart}

\RequirePackage{amsthm,amsmath,amsfonts,amssymb,dsfont}
\RequirePackage[authoryear]{natbib}
\RequirePackage[colorlinks,citecolor=blue,urlcolor=blue]{hyperref}
\RequirePackage{graphicx}

\startlocaldefs
\usepackage[dvipsnames]{xcolor}
\usepackage{tikz}
\usetikzlibrary{arrows.meta, positioning}

\usepackage{bm}
\usepackage{paralist}

\makeatletter 
\def\journal@name{}
\makeatother

\theoremstyle{plain}
\newtheorem{theorem}{Theorem}

\newcommand{\Exp}{\mathrm{E}}
\newcommand{\Cov}{\mathrm{Cov}}
\newcommand{\Var}{\mathrm{Var}}

\newcommand{\R}{\mathbb R}
\newcommand{\N}{\mathbb N}
\newcommand{\diff}{\mathrm d}
\newcommand{\eps}{{\varepsilon}}
\newcommand{\indicator}{\mathds{1}}

\endlocaldefs

\begin{document}

\begin{frontmatter}
\title{Copulas for Geostatistical Data: Foundations, Modeling Principles and Statistical Inference}
\runtitle{Copulas for Geostatistical Data}

\begin{aug}
\author[A]{\fnms{Axel}~\snm{B\"ucher}\ead[label=e1]{axel.buecher@rub.de}\orcid{0000-0002-1947-1617}}
\and
\author[B]{\fnms{Nadja}~\snm{Klein}\ead[label=e2]{nadja.klein@kit.edu}\orcid{0000-0002-5072-5347}}
\address[A]{Axel Bücher is Professor at Ruhr-Universität Bochum, Fakultät für Mathematik, Bochum, Germany\printead[presep={\ }]{e1}.}
\address[B]{Nadja Klein is Professor at Karlsruhe Institute of Technology, Scientific Computing Center, Zirkel 2, 76131 Karlsruhe, Germany\printead[presep={\ }]{e2}.}
\end{aug}

\begin{abstract} 
Spatial statistics commonly describes spatial dependence through second-order quantities such as covariance functions and variograms, often within Gaussian random-field models and under structural assumptions such as stationarity, isotropy, or distance-based decay. Copulas offer a complementary framework that separates marginal distributions from dependence and permits a broad range of non-Gaussian dependence structures. Because the finite-dimensional distributions of a spatial random field can always be decomposed into margins and copulas through Sklar's theorem, copulas provide a natural language for studying spatial dependence beyond second-order summaries. Yet the relevant literature has developed along several largely separate strands across spatial statistics, copula modeling, stochastic processes, and application domains, often with different terminology and modeling objectives. This review brings these strands together: We revisit classical concepts from spatial statistics through a copula lens, discuss copula-based tools for describing spatial dependence, and systematically review constructions of spatial copula models. Particular emphasis is placed on Kolmogorov consistency and on the distinction between models defined for a fixed set of locations and genuinely process-level constructions. We also discuss statistical inference, extensions to spatio-temporal settings, and emerging directions involving flexible marginal and dependence models. By clarifying the relationships among existing approaches and their respective strengths and limitations, the review provides a unified perspective on the interface between copula modeling and spatial statistics.
\end{abstract}


\end{frontmatter}



\section{Introduction}
\label{sec:introduction}

Spatial statistics concerns the analysis of phenomena indexed by geographical location and, more generally, by spatial domains. Its central objective is to describe, estimate, and predict spatial variation while accounting for dependence between observations collected at different locations. In many applications, systematic variation is first modeled through observed covariates, while a spatial random field is used to capture the remaining dependence. 

Spatial dependence, whether considered marginally or conditionally on covariates, is a defining feature of spatial data. Its modeling is often guided by Tobler's first law of geography, according to which nearby phenomena tend to be more strongly related than distant ones \citep[p.~236]{Tobler1970}. Classical geostatistics typically characterizes such dependence through second-order quantities, most notably covariance functions and variograms, and often embeds these within a Gaussian random-field framework \citep{Cressie1993}. Central to this framework are structural concepts such as stationarity and isotropy, which describe how the distributional or second-order properties of a spatial process behave under translations and rotations of the spatial domain. Their prominent role partly reflects a distinctive feature of spatial statistics: inference is often based on a single realization of the process observed at a finite collection of locations, so that meaningful inference about its dependence structure requires assumptions that permit information to be pooled across space.

A second conceptual foundation of this review is provided by copula theory, which offers a general framework for separating marginal distributions from the dependence structure. By Sklar's theorem, any multivariate distribution can be represented in terms of its univariate margins and a copula; when the margins are continuous, this copula is unique and fully characterizes the dependence structure \citep{Nelsen2006}. Irrespective of uniqueness, this representation makes it possible to specify marginal models and a joint dependence model as distinct components, and it has led to a vast literature on flexible multivariate modeling, also known as copula modeling \citep{Joe2015}. Copula families can accommodate a wide range of dependence features, including nonlinear association, tail dependence, and asymmetry. These features are not exclusive to copula models, but the copula framework provides a systematic and interpretable way to model them separately from the margins, particularly when Gaussian or, more generally, elliptical dependence structures are too restrictive.

Given these backgrounds, the combination of copula methodology and spatial statistics seems natural. Indeed, as we will see in Section~\ref{subsec:spatial-random-fields}, the principal stochastic building blocks of geostatistics, and of much of spatial statistics more broadly, are spatial random fields, whose distributions are characterized by their finite-di\-mensional distributions. By Sklar’s theorem, each finite-dimensional distribution admits a copula representation. Consequently, every spatial random field induces a family of finite-di\-mensional copulas that describes its dependence structure separately from its marginal distributions. This observation suggests several complementary perspectives. One may study the copulas induced by established random-field models, use copula-based quantities to characterize spatial dependence, or ask whether a prescribed family of copulas can serve as the basis for constructing a coherent spatial random field. Together with the associated questions of modeling and inference, these perspectives form the central focus of this review.

A first intuition might be that one can simply replace Gaussian finite-dimensional distributions by more flexible copulas. However, unlike ordinary multivariate modeling, the constructive perspective---that of defining a spatial random field from a prescribed copula family---raises challenges that do not arise in classical fixed-dimensional copula modeling. At the process level, the relevant object is a copula random field, that is, a random field with uniform marginal distributions. Such a field is determined not by a single multivariate copula, but by an entire family of finite-dimensional copulas indexed by arbitrary collections of spatial locations. For this family to define a coherent stochastic process, its members must be compatible under permutations and marginalization and hence satisfy the relevant Kolmogorov consistency conditions. These requirements substantially restrict the copula constructions that can be transferred from fixed-dimensional settings to spatial domains. In particular, the availability of a valid copula model in every fixed dimension does not by itself ensure that spatially indexed versions of these copulas form a coherent random field, especially when their parameters depend on the spatial configuration. The distinction between models defined only for a fixed set of locations and genuinely process-level constructions is therefore fundamental to the development, interpretation, and assessment of spatial copula models.

Despite these challenges, copula models have been applied to a broad range of problems involving geostatistical data, starting with the seminal contribution of \citet{Bardossy2006} and extending to spatial extremes \citep{DavisonHuser2014}, spatial time series \citep{ErhCzaSch2015}, replicated spatial measurements \citep{Krupskii2018}, and a variety of further applications \citep[see, e.g.,][]{Philippe2020,Huang2022}.
However, the resulting literature has developed along several largely separate strands. Some contributions employ finite-dimensional copula models tailored to a fixed collection of locations, whereas others consider copulas induced by established random-field constructions or develop genuinely process-level models. These contributions are dispersed across the literature on copulas, spatial statistics, stochastic processes, and applications, often using different terminology and pursuing different modeling objectives. Consequently, the relationships between the various approaches---and, in particular, the distinction between finite-domain models and coherent copula random fields---are not always made explicit. To the best of our knowledge, no existing review systematically brings these strands together despite addressing closely related questions.

Against this background, the present review seeks to provide a unified perspective on the different strands of this literature. We first revisit classical tools from spatial statistics through a copula lens and discuss copula-based notions of spatial dependence. 
We then systematically review existing approaches to constructing copula models for spatial random fields, with particular emphasis on the distinction between finite-domain and process-level constructions---a central theme of this review---and the role of Kolmogorov consistency. 
Further topics include statistical inference for spatial copula models, extensions to spatio-temporal settings, and emerging research directions. Throughout, our goal is to clarify the relationships between existing approaches, assess their respective strengths and limitations, and establish a common perspective for further developments at the interface of copula modeling and spatial statistics.

The remainder of the paper is organized accordingly. Section~\ref{sec:general-overview} introduces, in two separate parts, the key concepts from spatial statistics and copula theory that are needed throughout the review. From Section~\ref{sec:copula-tools-for-spatial-data} onward, the two fields are brought together. Section~\ref{sec:copula-tools-for-spatial-data} examines how copula-based tools can complement classical concepts to describe and analyze spatial dependence. Section~\ref{sec:copula-random-fields} reviews copula constructions for spatial random fields, while the associated inferential methods are discussed in Section~\ref{sec:statistics-spatial-copulas}. Section~\ref{sec:temporal-dependence} considers extensions to spatio-temporal settings. Finally, Section~\ref{sec:discussion-outlook} summarizes the main insights, identifies open problems, and outlines emerging directions for future research.

\section{Key concepts: Background and preliminaries}
\label{sec:general-overview}

In this section, we provide a brief overview of key concepts, tools, and models from spatial statistics (Section~\ref{sec:spatial-statistics-overview}) and from the theory and statistics of copulas (Section~\ref{sec:statistics-for-copulas-overview}). This overview establishes the foundations for connecting the two fields in the subsequent sections.

\subsection{Spatial statistics} 
\label{sec:spatial-statistics-overview}

\subsubsection{Spatial data structures} 

A common classification in spatial statistics is based on the underlying data structure \citep[Chapter 1]{Cressie1993}, thereby distinguishing between  \emph{point pattern data},  \emph{lattice data}, and \emph{geostatistical data.}

\emph{Point pattern data} arise when the observations are event locations, such as sites where precipitation exceeds a given threshold. 
A primary goal with point pattern data is to understand how points are distributed in space, and modeling is typically based on spatial point processes, with the Poisson process serving as the fundamental building block.
First‑ and second‑order characteristics, including the intensity and pair correlation functions, are established tools for describing spatial structure in this field \citep{Dig2014}.

\emph{Lattice (or areal) data} consist of observations aggrega\-ted over spatial regions such as counties or districts, for example disease rates per administrative unit. 
These data are typically modeled using lattice-based approaches, such as Markov random fields  \citep[MRFs; ][]{RueHel2005}. 
Prominent examples include the conditional autoregressive (CAR) models \citep{Bes1974}, intrinsic CAR models \citep{BesYorMol1991} which are used in Bayesian spatial models, Gaussian MRFs which yield sparse precision matrices, or simultaneous autoregressive models \citep{Cressie1993}. 
However, also copulas have been considered for this type of data \citep{Musgrove2016}.

Finally, \emph{geostatistical data} comprise measurements of a continuously varying spatial phenomenon observed at a finite set of locations within a continuous spatial domain, for instance wind speed at wind turbine sites. The standard modeling framework here is that of spatial random fields, where the underlying spatial phenomenon is treated as a realization of a stochastic process, most frequently a Gaussian process (GP).  This model-based geostatistical framework provides a coherent basis for interpolation \citep[Kriging;][]{Ste1999}, prediction, uncertainty quantification, and the incorporation of covariates or nonstationarity \citep[][]{Cressie1993,DigRib2007,BanCarGel2015}.

This review focuses on geostatistical data, and in particular on models based on spatial random fields. This focus is natural because, as explained in Section~\ref{sec:statistics-for-copulas-overview}, a basic prerequisite for copula methodology is the presence of random vectors. Such vectors arise naturally for geostatistical data by evaluating a random field at finitely many locations, and also for lattice data indexed by a finite set of spatial units.
However, the latter are often discrete, which creates additional complications for copula-based modeling. Point pattern data are different in nature: they do not directly give rise to random vectors indexed by fixed spatial locations and therefore fall outside the main scope of this review.

\subsubsection{Spatial random fields}
\label{subsec:spatial-random-fields}
A (univariate) \emph{spatial random field} is a  stochastic process $\mathbb Y = \lbrace Y(s)\rbrace_{s \in \mathcal S}$ indexed by some spatial domain $\mathcal S \subset \R^2$ or $\mathcal S \subset \R^3$. Mathematically, the stochastic properties of $\mathbb Y$ are characterized by its \emph{finite-dimensional distributions (fidis)}, that is, by the collection $\{F_{s_1, \dots, s_d}: s_1, \dots, s_d \in \mathcal S \text{ pairwise distinct}, \allowbreak d \in \N \}$ of cumulative distribution functions (cdfs), where 
\begin{multline*}
F_{s_1, \dots, s_d}(y) 
= 
\Pr\big[ Y(s_1) \le y_1, \dots, Y(s_d) \le y_d \big] , 
\\ 
y =(y_1, \dots, y_d)^\top \in \R^d.
\end{multline*}
These fidis encode the joint behavior of the field at any finite collection of locations---for example, soil-moisture measurements at two nearby sites or temperature readings at three weather stations.

\subsubsection{Structural assumptions}
\label{subsubsec:structural-assumptions}
Classical methods in spatial statistics typically do not rely on a full specification of the fidis (and hence of the spatial random field), but
rather rely on second-order characteristics, that is, quantities that depend only on the first and second moments of the field. These include the mean function, the covariance function, or the \emph{semivariogram}, defined for $s_1, s_2 \in \mathcal S$ as
\begin{align*}
\mu(s_1) &:= \Exp\big[Y(s_1)\big], \\ 
\Cov(s_1, s_2) &:= \Exp\big[(Y(s_1) - \mu(s_1))(Y(s_2) - \mu(s_2))\big], \\
\gamma(s_1, s_2) &:= \frac12\Var\big[ Y(s_1) - Y(s_2)\big ],
\end{align*}
respectively. 
To make these quantities tractable, one typically imposes additional structural assumptions. 
\emph{Second-order stationarity} refers to the assumption that the mean is constant, $\mu(s) \equiv \mu$, and that the covariance depends only on the spatial lag $h=s_1 - s_2$, that is, $\Cov(s_1, s_2) = K(h)$ for a function $K$ defined on the set of  differences between two elements  $s_1,s_2\in\mathcal S$.
Under this assumption, the semivariogram likewise depends only on the lag, i.e., $\gamma(s_1,s_2)=\gamma(h)$. In practice, second-order stationarity is often assumed only after a suitable transformation of the spatial field, for instance after removing a non-constant mean function or adjusting for covariate effects.

A further common simplification is \emph{isotropy}, which means that spatial dependence (measured by covariances) depends only on the Euclidean distance $\| h \|$, but not on the direction. In this case, $K(h) = K(\| h \|)$ and $\gamma(h) = \gamma(\| h \|)$.
A widely used isotropic covariance family in geostatistics 
is the \emph{Mat{\'e}rn class}.  For a spatial lag vector $h = s_1 - s_2$, the Mat{\'e}rn 
covariance function is given by
\[
K_{\text{Mat{\'e}rn}}(h)
= \sigma^2\,\frac{2^{1-\nu}}{\Gamma(\nu)}
(\kappa \|h\|)^{\nu} K_{\nu}(\kappa \|h\|),
\]
where $\sigma^2>0$ is the variance, $\kappa>0$ a scale (inverse range) parameter, $\nu>0$ the smoothness controlling the differentiability of the field, and $K_\nu$ the modified Bessel function of the second kind. Under a suitable rescaling of the range parameter, the Mat{\'e}rn covariance converges to the squared exponential model as $\nu \to \infty$, while for $\nu = 1/2$ it reduces to the exponential covariance. We also refer to  \citet{LinRueLin2011} for an explicit link between Gaussian fields and Gaussian Markov random fields \citep[GMRFs;][]{RueHel2005} via a stochastic partial differential equation (SPDE) to allow for non-stationarity.

If spatial dependence varies with direction, the field is \emph{anisotropic.} A common parametric approach assumes geometric anisotropy, which is based on replacing the Euclidean norm by a transformed distance, $\| h \|_A = \| A h \|_2$, where $A$ is a positive definite matrix in $\R^{2 \times 2}$ encoding rotation and axis-specific scaling. More generally, covariance models are often constructed as parametric functions of (possibly warped) domain distances,
\[
\Cov(s_1, s_2) = K_{\theta}\bigl(d_{\psi}(s_1, s_2)\bigr),
\]
where $d_\psi: \mathcal S \times \mathcal S \to [0,\infty)$ is a suitably chosen metric or deformation of the spatial domain. 
Such distance-based modeling provides a flexible yet structured framework that underlies most parametric covariance families used in geostatistics and forms the basis for many standard methods.

\subsubsection{Spatial observation schemes} 
\label{subsubsec:spatial-observation-schemes}
Before outlining key goals and methods in spatial statistics, it is important to address a fundamental feature concerning the availability of spatial data: on the one hand, in the classical setting, only a single realization of the spatial process is observed at a finite set of locations, denoted by $\mathcal S_d = \{ s_1, \dots, s_d \} \subset \mathcal S$. In this situation, structural assumptions such as (second-order) stationarity become indispensable for meaningful statistical inference. In addition, some form of ergodicity \citep[Section 2.3]{Cressie1993} is typically required to justify replacing population characteristics (e.g., means or covariances) by their empirical counterparts computed from a single spatial realization. 
On the other hand, structural assumptions remain useful, though not strictly necessary, in repeated-measurement settings. There, one observes at the same finite set  $\mathcal S_d$ a sample of $Y=(Y(s_1), \dots, Y(s_d))^\top$ of size $n$, say $Y_1, \dots, Y_n$. In most applications, these replicates are collected over time, placing the problem in the domain of spatio-temporal statistics. Temporal dependence between observations may then arise, although in many cases the sample can reasonably be treated as independent. We return to temporally dependent settings in Section~\ref{sec:temporal-dependence}.

\subsubsection{Key goals and methods in spatial geostatistics}
\label{subsec:goals-spatial-geostatistics}
We end this section by describing some of the key goals and methods in spatial statistics for geostatistical data, with details described in \cite{Cressie1993, DigRib2007, BanCarGel2015}, among others. A fundamental task is the characterization and estimation of the spatial dependence structure of the process. This is typically achieved through covariance functions and variograms (Section~\ref{subsubsec:structural-assumptions}), with empirical 
variograms providing an exploratory tool for assessing spatial dependence, while parametric covariance or variogram models are commonly fitted to obtain valid dependence structures that can subsequently be used for prediction and simulation.

A closely related objective is \emph{spatial prediction} (or \emph{spatial interpolation}/\emph{extrapolation}), where the aim is to predict the value of the spatial process at unobserved locations. The classical and most widely used method for this task is \emph{kriging}, which yields the best linear unbiased predictor 
under a specified mean structure and covariance or variogram model.
Kriging can also be interpreted as posterior mean prediction under a GP model \citep{RasWil2006}.

Finally, spatial statistics frequently addresses \emph{spatial regression} problems, where the goal is to explain variation in the spatial process through covariates while accounting for residual spatial dependence. Such models combine regression components with spatially correlated error terms, often represented through GPs or related random field models.

\subsection{Statistics based on copulas} 
\label{sec:statistics-for-copulas-overview}

\subsubsection{Copulas and Sklar's theorem}
A real-valued function $C$ on $\R^d$ is called a \emph{copula} if it is the cdf of some $d$-di\-mensional random vector with standard uniform marginal distributions. Since such a distribution is supported on the unit cube $[0,1]^d$, the values of $C$ outside this set are trivial. It is therefore customary to restrict $C$ to the unit cube, and to regard it as a function $C:[0,1]^d \to [0,1]$.

\begin{theorem}[Sklar's theorem]
\label{theo:sklar}
Let $Y=(Y_1,\dots,Y_d)$ be a $d$-dimensional random vector with joint cdf $F$ and marginal cdfs $F_1,\dots,F_d$. Then there exists a copula $C$ such that, for all $y\in\R^d$,
\begin{align} \label{eq:sklar-identity}
F(y)=C\bigl(F_1(y_1),\dots,F_d(y_d)\bigr).
\end{align}
If the marginals $F_1,\dots,F_d$ are continuous, then $C$ is unique, and it can be written as $C(u) = \Pr(U \le u)$, where $U=(U_1, \dots, U_d)$ has coordinates $U_j = F_j(Y_j)$.
Otherwise, $C$ is uniquely determined on $F_1(\R)\times\cdots\times F_d(\R)$. Conversely, for any copula $C$ and any univariate cdfs $F_1,\dots,F_d$, the function $F$ defined by \eqref{eq:sklar-identity} is a $d$-dimensional distribution function with marginal cdfs $F_1,\dots,F_d$.
\end{theorem}

Sklar's theorem has three key implications for statistics: first, 
since the left-hand side of \eqref{eq:sklar-identity} fully characterizes the distribution and thus the dependence structure of~$Y$, the same must be true for the right-hand side. However, the marginal cdfs do not carry any information about the dependence, whence all aspects of dependence must be encoded in the copula $C$.
Second, the uniqueness statement in Sklar's theorem is fundamentally an identifiability result: in a fully nonparametric setting, the copula is identifiable if and only if all marginal cdfs are continuous, and then it indeed \emph{characterizes} dependence.
Finally, the converse direction of Sklar's theorem is most important for modeling: the distribution of any random vector can be modelled by specifying a copula and by specifying marginal cdfs, even when the marginals are not continuous.

\subsubsection{Dependence measures} 
As explained in the previous section, copulas provide a natural framework for studying the dependence structure of multivariate distributions. In particular, classical bivariate dependence measures beyond the simple Pearson correlation, such as Kendall's $\tau$ or Spearman's $\rho$, can be expressed as functionals of the copula \citep{SchSchBluGaiRup2010}; see also Section~\ref{sec:Kendall-Spearman-variograms} below. The copula further captures qualitative features of dependence, such as symmetry and tail behavior. \emph{Tail dependence} refers to the tendency of random variables to simultaneously attain extreme values. In the bivariate case, this phenomenon is commonly quantified through the \emph{lower and upper tail dependence coefficients} $\lambda_L,\lambda_U \in [0,1]$ \citep{EmbMcnSta2002}. For a random vector $(Y_1,Y_2)$ with marginal cdfs $F_1$ and $F_2$ and copula $C$, these coefficients are defined by
\begin{align*}
\lambda_L &= \lim_{u \downarrow 0} \Pr(F_1(Y_1) \le u \mid F_2(Y_2) \le u) =  \lim_{u \downarrow 0} \frac{C(u,u)}{u},\\
\lambda_U
&=\lim_{u \uparrow 1}
\Pr(
F_1(Y_1) > u
\mid 
F_2(Y_2) > u)\\&=
\lim_{u \uparrow 1}
\frac{1 - 2u + C(u,u)}{1-u},
\end{align*}
provided the respective limits exist.
A positive value of $\lambda_U$ indicates asymptotic dependence in the upper tail, whereas $\lambda_U=0$ corresponds to asymptotic independence in the upper tail; analogous interpretations hold for $\lambda_L$.

\subsubsection{Constructing copulas}\label{subsubsec:construct}

There is a wide range of methods for constructing parametric copulas \citep{Nelsen2006, Joe2015}, and we can only discuss the most important ones. A particularly simple and versatile approach, which will be central to Section~\ref{sec:copula-random-fields}, is based on recovering the copula (or dependence structure) from a given joint distribution. The approach is also known as \emph{inverting Sklar's theorem}, and the resulting copulas are commonly referred to as \emph{inversion copulas} or \emph{implicit copulas} \citep{Nelsen2006, Smith2023}. The canonical example is the Gaussian copula. Start with a multivariate Gaussian cdf $F=\Phi_\Sigma$ with positive semidefinite correlation matrix  $\Sigma$ and standard Gaussian marginal cdfs $F_j=\Phi$. By Sklar's theorem, there exists a unique copula that couples the marginals $F_j$ to $F$, namely the Gaussian copula, given by $C_{\Sigma}(u) = \Phi_{\Sigma}(\Phi^{-1}(u_{1}),\dots,\Phi^{-1}(u_{d}))$. 
The same construction principle yields further important copula families. For instance, starting from a multivariate $t$-distribution produces the \emph{$t$-copula}, which exhibits tail dependence. More generally, elliptical reference distributions give rise to the broader class of \emph{elliptical copulas} \citep{FrahmJunkerSzimayer2003}. Likewise, applying this approach to max-stable distributions yields certain \emph{extreme-value copulas}, which play a central role in multivariate and spatial extreme-value modeling, including the Brown–Resnick model \citep{HusRei1989,KabSchHaa2009}, the Schlather model \citep{Sch2002}, and the extremal‑t process \citep{Opi2013}; see also \citet{DavPadRib2012} for a review.

Extreme-value copulas can also be constructed  from lower-dimensional functions, namely, from convex and homogeneous functions $\ell:[0,\infty)^d \to [0,\infty)$ called \emph{stable tail dependence functions}; see \cite{GudendorfSegers2010} and \cite{Ressel2013} for parametric families and characterizations of $\ell$. The canonical example from this class is the Gumbel-Hougaard copula. Similar to extreme-value copulas, \emph{Archimedean copulas} are constructed from functions $\psi:[0,\infty] \to [0,1]$ satisfying $\psi(0)=1$ and $\psi(\infty)=0$ which are strictly decreasing on $[0, \inf\{x: \psi(x)=0\}]$ and $d$-monotone \citep{McNeilNeslehova2009}, with a prominent example being the Clayton copula.
The classes of Archimedean and extreme-value copulas can, in fact, be unified within the broader class of \emph{Archimax copulas}; see \cite{CharpentierFougeresGenestNeslehova2014}.

Hierarchical Archimedean copulas \citep[HACs;][]{OkhOkhSch2013} provide a parsimonious multivariate extension of Archi\-medean copulas by nesting generators according to a tree structure. While this induces flexible clustered dependence, the resulting copulas remain exchangeable within subclusters and inherit symmetric tail behavior from their
generators. In addition, the nesting construction is subject to nontrivial compatibility conditions on the generators to ensure that the resulting copula is well defined. These limitations motivate more flexible constructions that are particularly useful in higher dimensions, such as vine or factor copulas.

\emph{Vine copulas} are based on decomposing a multivariate copula into a sequence of bivariate copulas (also known as a pair-copula construction) organized through a graphical structure called a vine \citep{AasCzadoFrigessi2009}. More specifically, the joint copula is expressed as a product of conditional bivariate copulas arranged across several linked trees. This allows different copula families to be used for different pairs of variables, enabling highly flexible dependence modeling in high dimensions \citep{Cza2019, CzaNag2022}.

\emph{Factor copulas} are based on introducing one or several latent variables (factors) that drive the dependence among the observed variables. In these models, the components of the random vector are typically assumed to be conditionally independent given the latent factors, while the unconditional dependence arises through their common exposure to these factors. This leads to parsimonious dependence structures that remain tractable even in high dimensions, while still allowing for flexible tail behavior and asymmetric dependence \citep{KruJoe2013, OhPat2017}; see also Section~\ref{subsubsec:factor-copula-fields}.

Finally, many methods allow one to generate \emph{new copulas from old}, for instance based on rotations or reflections of the copula distribution or on mixtures. We refer to \cite{Nelsen2006} for an overview in the bivariate case. A construction particularly useful for spatial application arises from non-monotone transformations of the margins of a copula vector $U$ with cdf $C$ \citep{Quessy2024}; this will be discussed in more detail in Section~\ref{subsubsection:new-from-old-via-non-monotone-mappings}.

\subsubsection{Key goals and methods in statistics for copulas}
\label{subsubsec:key-goals-methods-statistics-copulas}

The main statistical goal of copula modeling is to describe and analyze the joint distribution of a random vector $Y=(Y_1,\dots,Y_d)$ by specifying an appropriate copula and the $d$ marginal distributions.  Once estimated, the resulting joint distribution can be used for a wide range of purposes, including simulation, prediction, conditional distributional analysis, risk assessment, and the study of tail dependence. In the presence of covariates, this naturally leads to a multivariate form of distributional regression \citep{Kle2024}, in which both the marginal distributions and the copula may depend on explanatory variables \citep[see, e.g.,][for recent contributions]{KleHotBarKne2022,KocKle2025}.

In the basic setting of an independent sample $Y_1,\dots, \allowbreak Y_n$ from $Y$, $Y_i=(Y_{i1}, \dots, Y_{id})^\top$, inference may proceed either by estimating the marginal and copula parameters jointly, or by using a two-step approach. In the two-step case, the marginal cdfs $F_j$, $j=1,\ldots,d$ are estimated first, either parametrically or nonparametrically, to obtain estimates $\hat F_j$. The copula can then be estimated in a second step based on the (approximate) copula data $\hat {U}_1, \dots, \hat {U}_n$, obtained by transforming the data to approximately standard uniform margins via the estimated probability integral transform, that is, $\hat U_{ij}=\hat F_j(Y_{ij})$. If the margins are estimated parametrically, the approach is referred to as \emph{inference functions for margins}
\citep{JoeXu1996}.

While joint estimation is statistically appealing, two-step approaches are often substantially easier to implement, particularly in high-dimensional models. Under suitable regularity conditions, two-step estimators retain good asymptotic properties, although the potential efficiency loss and the contribution of first-stage marginal estimation to overall uncertainty are model-dependent and need not be negligible \citep{Joe2005, GenGhoRiv1995, Tsu2005}. Recent work nevertheless suggests that the impact of marginal estimation on overall uncertainty can be small in a range of correctly specified and moderately misspecified models, including Bayesian formulations \citep{Smi2013, KoHjo2019, Smith2025}. In such settings, standard delta-method arguments imply that uncertainty assessments for smooth predictive functionals should also be similar, as observed in copula-based regression models \citep{KleKne2016, SmithKlein2021}.

Depending on the model class and dimension, the estimation steps may rely on standard statistical principles, including (penalized) maximum likelihood, composite likelihood, Bayesian methods, or other forms of regularization. 
For reviews and discussions of further methods and extensions to time series, count data, and covariates, we refer to \cite{GenNes2007, Smi2013, DurSem2015, Joe2015, CzaNag2022, Smith2023} and the references therein.

\section{Copula-based tools for spatial models} 
\label{sec:copula-tools-for-spatial-data}

Before turning to copula-based modeling of spatial data in the subsequent sections, we first discuss how the copula framework offers tools for spatial analysis, enabling the study of dependence features that go beyond traditional second-moment approaches. In doing so, we follow the historical development of the connection between the two fields, which began with the seminal contributions of \cite{Bardossy2006}.

The starting point is the study of bivariate spatial dependence under a suitable stationarity assumption that naturally extends the second-order stationarity from Section~\ref{sec:spatial-statistics-overview}. By Sklar’s theorem, for any pair  $(s_1, s_2)$ of distinct locations in $\mathcal S$, the joint cdf $F_{s_1,s_2}$ of $(Y(s_1), Y(s_2))$ can be written as $F_{s_1,s_2}(y_1,y_2) = C_{s_1, s_2}(F_{s_1}(y_1), F_{s_2}(y_2))$ for some copula $C_{s_1,s_2}$. The assumption of \emph{strict bivariate stationarity} requires the bivariate fidis of $\mathbb Y$ to satisfy $F_{s_1, s_2}=F_{s_1+h,s_2+h}$ for every admissible spatial translation $h \in \R^2$. Under this assumption, the marginal cdfs are spatially invariant, $F_{s_j} \equiv F_Y$, and if $F_Y$ is additionally continuous, the bivariate copulas $C_{s_1, s_2}$ depend only on the spatial lag $h=s_1 - s_2$: 
\begin{align*}
C_{s_1, s_2}(u_1, u_2) = C_h(u_1, u_2), 
\quad 
(u_1, u_2) \in [0,1]^2,
\end{align*}
for some bivariate copula $C_h$ indexed by $h$. Note that $C_h(u_1, u_2) = C_{-h}(u_2, u_1)$.

\subsection{Copula-based variograms for spatial data}

The starting point for copula-based variograms is the \emph{indicator semivariogram} at cutoff level $\beta\in \R$ \citep[p.~282][]{Cressie1993}, defined by
\begin{align*}
\gamma_\beta(h) 
&=
\frac12 \mathbb E\Big[ \Big\{ \indicator(Y(s+h) \le \beta) - \indicator(Y(s) \le \beta) \Big\}^2 \Big].
\end{align*}
Note that  the right-hand side does not depend on $s$ under strict bivariate stationarity, which we implicitly assume throughout. If $F_Y$ is continuous, a straightforward calculation shows that $\gamma_\beta$ can be written in terms of the copula and the marginal cdf $F_Y$ as
\begin{align*}
\gamma_\beta(h) 
&=
F_Y(\beta) - C_h(F_Y(\beta), F_Y(\beta)),
\end{align*}
which is a correction of Equation (10) in \cite{Bardossy2006}.
The previous formula naturally suggests a margin-free measure of spatial dependence, which we call the \emph{copula semivariogram} 
\begin{align*}
\gamma_u^C(h) 
&=
u - C_h(u,u), \qquad u \in [0,1].
\end{align*}
Since  $u=C_\wedge(u,u)$, where $C_\wedge(u_1,u_2)=\min(u_1,u_2)$ is the upper Fréchet-Hoeffding bound (the copula corresponding to perfect positive dependence), it is essentially a measure of the distance between the upper Fréchet-Hoeffding bound and the spatial lag-$h$ copula $C_h$ along the main diagonal of the unit square. 
For $u=1/2$, we obtain that $\gamma_{1/2}^C(h) = (1-\beta_h)/4$, where $\beta_h$ denotes Blomqvist's beta of $(Y(s+h),Y(s))$ \citep{Blomqvist1950}.

A natural extension of the copula semivariogram is provided by the \emph{copula cross-semivariogram} defined, for $(u_1, u_2) \in [0,1]^2$ by
\begin{align*}
\gamma_{u_1, u_2}^C(h) 
&=
\frac12 \mathbb E\Big[ \Big\{ \indicator(Y(s+h) \le \beta_1) - \indicator(Y(s) \le \beta_1) \Big\}
\\& \qquad \times \Big\{ \indicator(Y(s+h) \le \beta_2) - \indicator(Y(s) \le \beta_2) \Big\}  \Big]
\\&= 
u_1 \wedge u_2 - \frac12 \big[ C_h(u_1, u_2) + C_{h}(u_2, u_1) \big],
\end{align*}
where $\beta_j = F_Y^{-1}(u_j)$.
Thus, $\gamma_{u_1,u_2}^C(h)$ measures the deviation of the upper Fréchet–Hoeffding bound from the symmetrized lag-$h$ copula, that is, the mixture copula $(C_h+C_{-h})/2$. Under isotropy, this mixture reduces to $C_h=C_{\|h\|}$. There is a clear parallel to copula- and quantile-based cross-covariance kernels in time series analysis; see, for instance, \cite{BarunikKley2019}, which rely on extending classical cross-covariances to  margin-free, copula-based measures. Finally, note that $\gamma_u^C(h)=\gamma_{u,u}^C(h)$.

\begin{figure}
    \centering
    \includegraphics[width=0.95\linewidth]{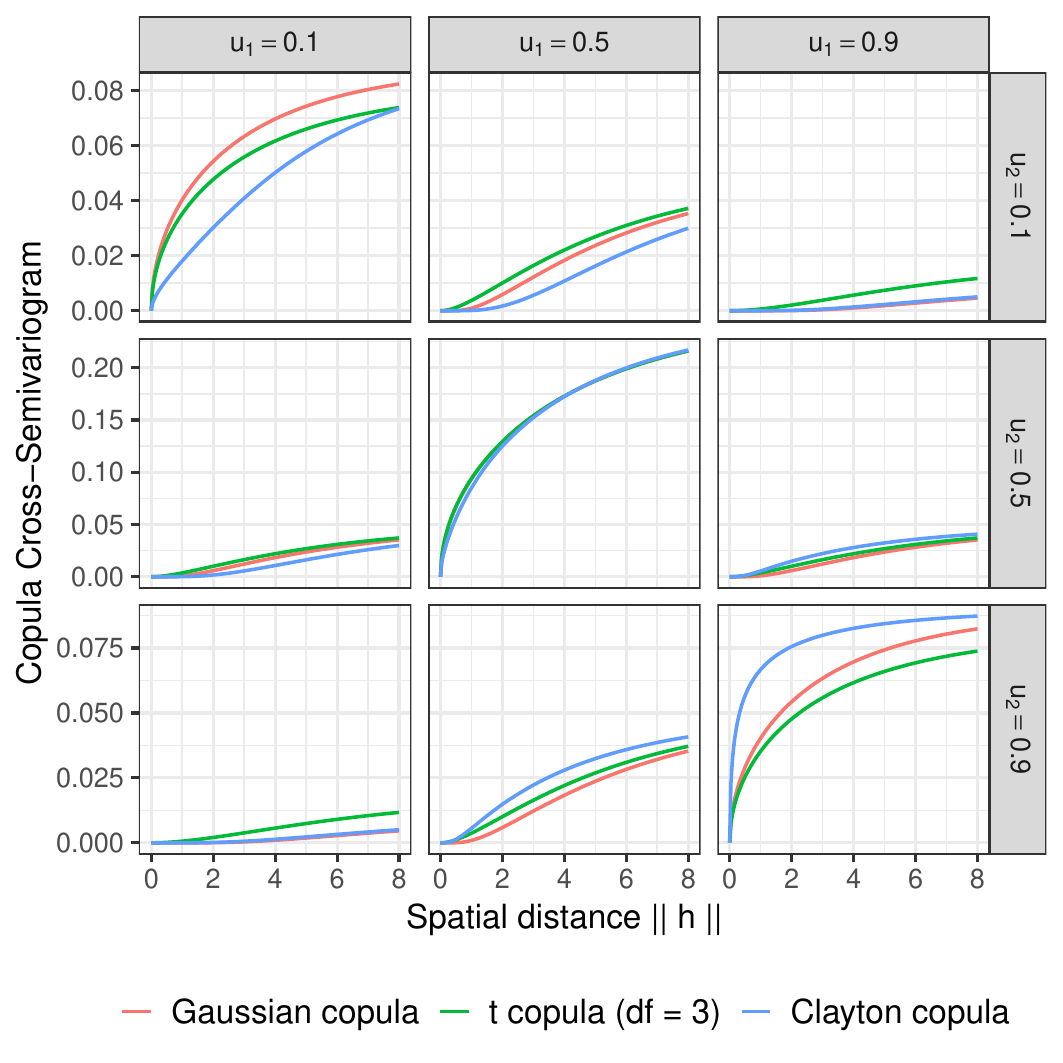} \vspace{-.2cm}
    \caption{Copula cross-semivariograms $\gamma_{u_1, u_2}^C(h) $ for the Gaussian, the $t_3$, and the Clayton copula, with $u_j \in \{.1,.5,.9\}$. The parametrization of each copula is chosen in such a way that the Spearman semivariogram from \eqref{eq:spearman-variogram} is $\gamma_S(h)=1-\exp(-\|h\|/5)$. Note that the Gaussian and $t$-copulas are symmetric with respect to the lower and upper tails, while the Clayton copula is asymmetric. Further, the Gaussian copula is tail independent in both tails, the $t_3$-copula is tail dependent in both tails, and the Clayton copula is tail dependent in the lower tail only.
    }
    \label{fig:variograms2}
\end{figure}

Figure~\ref{fig:variograms2} highlights the key advantages of copula-based dependence diagnostics over classical second-order tools. First, they capture central dependence through the behaviour of $\gamma_{u}^C(h)$ for intermediate values of $u$ (central panel). They also reveal asymmetries in the dependence structure by comparing $\gamma_{u}^C(h)$ with $\gamma_{1-u}^C(h)$ for  $u<1/2$ (radial asymmetry; upper left versus lower right panel; a feature of wind speed data, for instance), or by comparing $\gamma_{u,v}^C(h)$ with $\gamma_{v,u}^C(h)$ (exchangeability; upper right versus lower left panel). Finally, they provide insight into tail dependence, which can be assessed by contrasting $\gamma_{u}^C(h)$ with the corresponding Gaussian benchmark $\gamma_{u}^{C_{\mathrm{Gaussian}}}(h)$ for small or large values of $u$ (again visible in the upper left and lower right panels).

\subsection{Estimating copula-based variograms}
\label{subsec:estimating-copula-variograms}

Copula cross-semivariograms can be estimated by natural nonparametric analogues, both in the presence of repeated measurements and when only a single realization of $\mathbb Y$ is observed \citep{Bardossy2006}. 
For brevity, we focus on the latter case, where the available data are $Y(s_1), \dots, Y(s_d)$. First, the marginal cdf $F_Y$ can be estimated by the empirical cdf $\hat F_Y (y) = d^{-1} \sum_{i \in [d]} \indicator(Y(s_i) \le y)$, which gives rise to the  (approximate) copula data $\hat U(s_j) = \hat F_Y(Y(s_j))$; see Section~\ref{subsubsec:key-goals-methods-statistics-copulas}.
For a spatial bandwidth $b \ge 0$, the empirical copula cross-semivario\-gram is defined as
\[
\hat \gamma_{u,v}^C(h) = u \wedge v - \frac12\big[ \hat C_h(u, v) + \hat C_{h}(v, u) \big],
\]
where $\hat C_h$ denotes a spatially localized version of the empirical copula \citep{Deheuvels1979} defined by 
\begin{align*}
\hat C_h(u, v) 
= \frac1{|S_b(h)|} \sum_{(s, s') \in S_b(h)} \indicator\big[  
&\hat U(s) \le u, \hat U(s') \le v \big] ,
\end{align*}
with $S_b(h) = \{ (s, s') \in \mathcal S_d^2: s \ne s', \|  h- (s - s') \|_2 \le b \}$ assumed non-empty. Note that $S_b(h)$ collects all pairs of locations whose spatial lag has distance at most $b$ from $h$, and that $\hat U(s_j)=R(s_j)/d$ with $R(s_j)$ the rank of $Y(s)$ among $Y(s_1), \dots, Y(s_d)$; the empirical copula cross-semi\-vario\-gram is therefore rank-based.
An alternative rank-based estimator is given by
\begin{align*}
\tilde \gamma_{u, v}^C(h) = &\, \frac1{2|S_b(h)|} \sum_{(s, s') \in S_b(h)} \\
&\Big\{ \indicator(\hat U(s) \le u) - \indicator(\hat U(s') \le u) \Big\} \\
\times  &\Big\{ \indicator(\hat U(s) \le v) - \indicator(\hat U(s') \le v) \Big\}.
\end{align*} 
Both estimators raise theoretical questions that differ from those encountered in the standard empirical-copula setting \citep{Segers2012, BucVol2013}. In particular, $\hat F_Y$ is estimated from the full dependent spatial sample, so that the resulting pseudo-observations $\hat U(s)$ inherit an additional dependence structure. Their asymptotic behavior may therefore depend on the spatial sampling regime, the bandwidth sequence, and suitable ergodicity or mixing conditions. To the best of our knowledge, the theoretical properties of the estimators defined above have not yet been studied.

\subsection{Variograms induced by margin-free dependence coefficients}
\label{sec:Kendall-Spearman-variograms}

We end this section by briefly introducing additional variogram-type spatial diagnostics derived from the mar\-gin-free dependence measures presented in Section~\ref{sec:statistics-for-copulas-overview}. These diagnostics are likewise copula-based and, unlike the classical variograms discussed in Section~\ref{sec:spatial-statistics-overview}, do not require the existence of finite second moments. Like classical variograms, they describe the \emph{average strength of dependence} between observations separated by a spatial lag; in particular, they are not as informative as the family of copula (cross-)semivariograms.

First, under bivariate strict stationarity, the population version of the Kendall semivariogram (or \emph{$\tau$-semivario\-gram}) is defined as the dependence-dissimilarity measure
$
\gamma_{\tau}(h) := 1 - \tau(h),
$
where $\tau(h)$ denotes Kendall's $\tau$-coefficient computed from $(Y(s+h), Y(s))$. In view of Theorem~5.1.3 in \cite{Nelsen2006}, we may write
\[
\gamma_{\tau}(h)
= 2 - 4\,\mathbb{E}_{(U,V) \sim C_h}\!\left[
    C_h\!\bigl(U,V\bigr)
  \right],
\]
where $C_h$ is the spatial lag-$h$ copula of $(Y(s+h), Y(s))$. Likewise, the Spearman semivariogram is $\gamma_{S}(h) := \allowbreak 1 - \rho_S(h)$, where $\rho_S(h)$ is the population version of Spearman's correlation coefficient of $(Y(s+h),Y(s))$. In view of Theorem~5.1.6 in \cite{Nelsen2006}, we have
\begin{align} \label{eq:spearman-variogram}
\gamma_S(h)
=
4 - 12\,\mathbb{E}_{(U,V) \sim C_h}\!\left[UV\right].
\end{align}
Estimation of both $\gamma_{\tau}$ and $\gamma_S$ is straightforward and analogous to the approach described in Section~\ref{subsec:estimating-copula-variograms}; details are omitted for brevity.

\section{Copula models for spatial random fields} 
\label{sec:copula-random-fields}

In this section, we consider copula-based models for a (univariate) spatial random field $\mathbb Y=\mathbb Y^{\raisebox{-.2ex}{$\scriptstyle(\mathrm{s})$}}=\{Y(s)\}_{s\in\mathcal S}$, with spatial domain $\mathcal S\subset\mathbb R^2$ or $\mathcal S\subset\mathbb R^3$. A fundamental distinction arises depending on whether $\mathcal S$ is finite or infinite.
If $\mathcal S$ is finite, then $\mathbb Y$ may simply be regarded as a $d$-dimensional random vector, where $d=|\mathcal S|$, so that classical copula theory and Sklar's theorem, including the use of the reverse direction described in Section~\ref{subsubsec:construct},  apply directly. In contrast, if $\mathcal S$ is infinite---and typically continuous---copula modeling becomes substantially more challenging, since one must specify an entire family of finite-dimensional copulas satisfying suitable consistency conditions.

To make this precise, let $\{F_{s_1,\dots,s_d} : d\in\mathbb N,\ s_1,\dots,s_d\in\mathcal S \text{ pairwise distinct}\}$ denote the finite-dimensional distributions (fidis) of $\mathbb Y$. By Sklar's theorem, for every choice of pairwise distinct locations $s_1,\dots,s_d$, there exists a copula $C_{s_1,\dots,s_d}$ such that, for any $y\in\mathbb R^d$,
\[
F_{s_1,\dots,s_d}(y)
=
C_{s_1,\dots,s_d}\bigl(F_{s_1}(y_1),\dots,F_{s_d}(y_d)\bigr).
\]
We refer to the collection $\{C_{s_1,\dots,s_d} : d\in\mathbb N,\ s_1,\dots,s_d\in\mathcal S \text{ pairwise distinct}\}$ as the \emph{copula finite-dimensional distributions} (or \emph{copula fidis}) of $\mathbb Y$. When marginal cdfs are continuous, these coincide with the fidis of the associated \emph{copula random field} $\mathbb U=(U(s))_{s\in\mathcal S}$, where $U(s)=F_s(Y(s))$ has a standard uniform distribution for every $s\in\mathcal S$.

Consequently, specifying a spatial copula model on an infinite spatial domain amounts to specifying a copula random field, or, equivalently, its full collection of copula fidis. However, as discussed in Section \ref{subsec:Kolmogorov-consistency} below, an arbitrary collection $\{C_{s_1, \dots, s_d}: s_1, \dots, s_d \in \mathcal S \text{ pairwise distinct}, d \in \N \}$ of copulas need not define a bona fide copula random field. For this reason, copula field models on infinite domains are usually based on explicit stochastic constructions derived from well-known random field models, such as Gaussian, Gamma, or max-stable random fields.  Following the terminology introduced in Section~\ref{subsubsec:construct}, this approach may be called \emph{implicit copula field modeling}; it is discussed in Section~\ref{subsec:top-down-copula-field}. By contrast, models that are not based on such stochastic constructions are typically restricted to finite domains, with vine copulas being a prominent example. These models are discussed in Section~\ref{subsec:bottom-up}. Finally, Section~\ref{subsec:copula-covariates} considers extensions of spatial copula modeling that incorporate additional covariates.

Finally, we recall how copulas enter a complete spatial model. By the reverse direction of Sklar's theorem, any copula random field $\mathbb U$, together with prescribed marginal cdfs $\{F_{s}\}_{s\in\mathcal S}$, induces a spatial random field via $Y(s)=F_s^{-1}(U(s))$. Copula-based spatial models therefore naturally separate marginal behavior at individual locations from the spatial dependence structure. Our focus is on the latter; for marginal modeling, we refer to the textbooks mentioned in Section~\ref{sec:spatial-statistics-overview}. The interaction between marginal and copula modeling in estimation and inference is briefly discussed in Section~\ref{sec:statistics-spatial-copulas}.

\subsection{Kolmogorov consistency}
\label{subsec:Kolmogorov-consistency}

Throughout, we consider the situation where $\mathcal S$ is an infinite spatial domain.
Kolmogorov’s extension theorem \citep[Section~36]{Billingsley1995} then implies that a given collection $\{C_{s_1, \dots, s_d}: s_1, \dots, s_d \in \mathcal S \text{ pairwise distinct}, d \in \N \}$ of copulas can arise as the copula fidis of a spatial random field if and only if it satisfies the following two consistency conditions:
\begin{compactenum}
    \item \textbf{Permutation invariance.} For all pairwise distinct locations $s_1, \dots, s_d \in \mathcal S$, $d \in \N$, all permutations $\pi:[d] \to [d]$, and all $u \in [0,1]^d$, we have 
    $
    C_{s_{\pi(1)}, \dots, s_{\pi(d)}}(u_{\pi(1)}, \dots, u_{\pi(d)}) = C_{s_1, \dots, s_d}(u_1, \dots, \allowbreak u_d).
    $ 
    \item \textbf{Marginal consistency.} For all pairwise distinct locations $s_1, \dots, s_d \in \mathcal S$, $d \in \N$, all $k \in [d]$, and all $u_1, \allowbreak \dots, u_k \in [0,1]$, we have
    $
    C_{s_{1}, \dots, s_k}(u_1, \dots, u_k) \allowbreak = C_{s_1, \dots, s_k, s_{k+1}, \dots, s_d}\allowbreak (u_1, \dots, u_k, 1, \dots, 1).
    $
\end{compactenum}
Directly specifying an entire collection of copula fidis and verifying these conditions is typically difficult. This helps explain why such bottom-up constructions have received little attention in the literature. 
Structural assumptions such as stationarity or isotropy do not substantially mitigate this problem. For example, under strict stationarity on a translation-invariant domain, modeling a copula indexed by $d$ locations $s_1,\dots,s_d$ can be reduced to modeling a copula indexed by the $d-1$ relative lags $s_1-s_d,\dots,s_{d-1}-s_d$. However, this reduction concerns only the spatial parametrization; the Kolmogorov consistency constraints across dimensions remain fully operative. In other words, reducing the effective modeling dimension from $d$ to $d-1$ does not simplify the problem of finding Kolmogorov-consistent copula fidis.

\subsection{Implicit copula random field models for arbitrary spatial domains}
\label{subsec:top-down-copula-field}

As explained above, copula models on arbitrary (potentially infinite) spatial domains are typically obtained from established random field models through marginal standardization; Kolmogorov consistency is then automatic. For instance, a common assumption is that a spatial field $\mathbb Y$ follows a Gaussian random field model after monotone transformation to Gaussian margins \citep{ChilesDelfiner2012,GelSch2016}. Equivalently, the copula fidis of $\mathbb Y$ form a consistent family of Gaussian copulas. Similarly, in spatial extremes for block maxima, margins are often standardized to a fixed extreme-value distribution, such as the Fréchet$(1)$ distribution, and the transformed field is assumed to be max-stable \citep{DavPadRib2012, DavisonHuser2014}. This induces a consistent family of extreme-value copulas. For threshold exceedances, Pareto random fields \citep{FerDeh14} play an analogous role; see \citet{DomLegOpi2024} for a recent review.

Beyond these examples, implicit copula modeling has been used systematically to construct new and flexible classes of copula random fields on potentially infinite spatial domains. We review three representative classes next.

\subsubsection{New from old via non-monotone mappings}
\label{subsubsection:new-from-old-via-non-monotone-mappings}
A first approach to constructing new copula random-field models is to apply a fixed (possibly parametrized) non-monotone transformation to the margins of a given random field model, with the transformation chosen in such a way that the resulting field has standard uniform margins \citep[\emph{new copulas from old}; see][for a recent review]{Quessy2024}. This idea originates from the work of \cite{Bardossy2006}, whose key construction involves squaring a Gaussian random field $\{Z(s)\}_{s\in\mathcal S}$ with standard normal margins (i.e., applying the map $z \mapsto z^2$ to each $Z(s)$), and then transforming the squared field to the uniform scale.  The resulting copula fidis are known as \emph{chi-square copulas}, and the resulting copula random field can be shown to admit the representation $U(s)=|1-2\Phi(Z(s))|$, where $\Phi$ denotes the standard normal cdf.

Early extensions and applications of this construction appear in \cite{Bardossy2008,BardossyPegram2009}, while more rigorous mathematical treatments are provided in \cite{QuessyRivestToupin2015,Quessy2019,Nas2020}. These extensions typically involve non-monotone transformations of the margins beyond the squaring map. A convenient feature of the resulting models is that they allow parametrization through a correlation matrix (when starting from a Gaussian or $t$ random field), which in turn enables distance-based parametrizations in terms of spatial lags, as described in Section~\ref{subsubsec:structural-assumptions}. When the construction starts from a $t$ random field, the resulting copula fidis are known as \emph{Fisher copulas} \citep{FavQueTou2018,Brunner2019}.

\subsubsection{Factor copula random fields}
\label{subsubsec:factor-copula-fields}
A second class is based on factor representations. In the simplest setting, one starts from a random field $\mathbb W = (W(s))_{s \in \mathcal S}$ defined by
\[
W(s) = Z(s) + V,
\]
where $\mathbb Z = (Z(s))_{s \in \mathcal S}$ is a Gaussian field with standard normal margins, and $V$ is a common factor independent of $\mathbb Z$ \citep{Krupskii2018}. If $V = |N|$, where $N$ is standard normal, then $\mathbb W$ is called a skew-Gaussian field. The associated copula random field, defined by $U(s) = F_W(W(s))$, is referred to as a \emph{factor copula random field}. Formally, the model is parameterized by the correlation function of the Gaussian field $\mathbb Z$ and the distribution function $F_V$ of $V$. Hence, this construction permits parametrization of the dependence structure in terms of spatial distances.  \citet{Krupskii2018} study several theoretical properties of the model. Among other results, they derive conditions on $F_V$ under which the copula fidis exhibit tail dependence and show that the model is typically not identifiable unless replicated data are available \citep[see also][for the special case of skew-Gaussian fields]{GentonZhang2012}. The framework has been extended to multivariate spatial processes \citep{Krupskii2019}, to non-stationary settings \citep{MondalKrupskiiGenton2024}, and to spatio-temporal data \citep[][see also Section~\ref{subsubsec:space-time-factor-copulas} below]{KruGen2017}, with applications reported in \cite{CasHus2020}, among others.

\subsubsection{Archimedean copula random fields}
\label{subsubsec:archimedean-copula-fields}
A third class was recently proposed by \citet{Bevilacqua2024}. It is based on stochastic constructions related to Archimedean copulas, focusing in particular on the Clayton family. Their approach is motivated by the classical representation of the $d$-variate Clayton copula as the distribution of
\[
 U = \bigl(\varphi_a(E_1/M_a), \dots, \varphi_a(E_d/M_a)\bigr),
\]
where $E_1,\dots,E_d$ are iid standard exponential variables, $M_a$ is an independent Gamma$(1/a,1)$ variable, and $\varphi_a(x)=(1+x)^{-1/a}$ denotes the Laplace transform of $M_a$. The proposed \emph{Clayton random field} replaces these iid ingredients by dependent Gamma random fields, defining
\[
U(s)=\varphi_{a}(E(s)/M_{a}(s)), \qquad s \in \mathcal S,
\]
where $E(s)=G_2(s)$ and $M_{a}(s)=G_{2a}(s)$ are independent Gamma random fields parametrized through spatial correlation functions. In contrast to the classical finite-dimensional Clayton construction, the variables $E(s)$ are no longer iid across locations, and the normalization is performed pointwise rather than through a common random factor. Consequently, the resulting copula fidis are generally not Clayton copulas themselves. Nevertheless, the construction yields a valid copula random field with flexible spatial dependence induced through the latent Gamma random fields. Moreover, the bivariate marginal copulas $(U(s_1), U(s_2))$ have an explicit representation in terms of the underlying spatial correlation function \cite[Theorem 3]{Bevilacqua2024}, which allows for composite likelihood-based estimation; see Section~\ref{sec:statistics-spatial-copulas} for details.

\subsection{Vine copula models for finite spatial domains}
\label{subsec:bottom-up}

As briefly explained in Section~\ref{subsubsec:construct}, vine copulas \citep{AasCzadoFrigessi2009} form a flexible class of multivariate copula models. Recall that they are based on decomposing a high-dimensional copula into a sequence of bivariate copulas organized by a graphical structure consisting of a collection of trees. Since this construction is not based on an underlying stochastic process, it does not directly extend to infinite spatial domains, where Kolmogorov consistency conditions must be satisfied. Nevertheless, genuine spatial features can be incorporated in various ways, making vine copulas a useful approach for modeling finite spatial domains.

The earliest contribution on spatial vine copulas is due to \cite{GralerPebesma2011}, who aim to construct flexible copula models whose bivariate dependence structure is isotropic in the sense that it depends only on pairwise spatial distances.
Let the spatial domain be given by $\mathcal S=\{s_0,\dots,s_d\}$, and write $Y_{0:d}=(Y(s_0),\dots,Y(s_d))^\top$. 
Following \cite{GralerPebesma2011}, the copula of $Y_{0:d}$, denoted by $C_{0:d}$, is assumed to be a C-vine copula rooted at node $0$. Its Lebesgue density therefore exists and admits the factorization
\begin{align*}
c_{0:d}&(u_0,\dots,u_d)
=
\Big[ \prod_{i=1}^d c_{0,i}(u_0, u_i) \Big]
\\
&\times 
\Big[ \prod_{k=1}^{d-1}
\;\prod_{i=k+1}^{d}
c_{k,i\,|\,0:(k-1)}
\big(
u_{k\,|\,0:(k-1)},
\,u_{i\,|\,0:(k-1)}
\big)\Big],
\end{align*}
where $c_{0,i}$ and $c_{k,i\,|\,0:(k-1)}$ are bivariate copula densities (to be specified by the statistician) and where the values $u_{k\,|\,0:(k-1)}$ and $u_{i\,|\,0:(k-1)}$ can be obtained recursively from those copulas and from $u_0, \dots, u_d$.

A key feature of the C-vine construction is that the bivariate copula of $(Y(s_0), Y(s_i))^\top$ is precisely $C_{0,i}$, the copula corresponding to the density $c_{0,i}$. This makes the isotropy assumption easy to incorporate: writing $h_i = \|s_0 - s_i\|_2$, isotropy implies that $C_{0,i} = C_{h_i}$, where $C_h$ is a bivariate copula parameterized by the spatial distance $h$. \cite{GralerPebesma2011} propose a specific parametric model for $C_h$ involving a cascade of convex combinations.
\cite{Graler2014} extends the above construction by also assuming that the higher-order pair-copula densities $c_{k,i\,|\,0:(k-1)}$ only depend on the spatial distances $\| s_k - s_i\|_2$.

Another approach to modeling spatial dependence using vine copulas was proposed by \cite{ErhCzaSch2015}, albeit in a spatio-temporal setting that we will revisit briefly in Section~\ref{sec:temporal-dependence}. Their spatial dependence model is based on (truncated) R-vine copulas, which are constructed through a general sequence of trees satisfying certain proximity conditions, thereby providing greater flexibility than the C-vine model described above. 
Once the tree structure has been selected (which in practice is done in a data-adaptive manner using the maximum-spanning-tree algorithm of \citealp{Dissmann2013}), the parameters of the bivariate copulas are modeled as functions of spatial distances and elevation differences, enabling prediction at unobserved locations after model fitting under additional structural assumptions on the extended R-vine incorporating the new location. Estimation is carried out sequentially by maximizing likelihood functions separately at each tree level.

\subsection{Incorporating covariates into spatial copula models}
\label{subsec:copula-covariates}

Geostatistical data are often accompanied by additional covariates, such as topographical, meteorological, temporal, or land-use variables. These may be global, taking the same value across all locations, which is primarily meaningful in repeated-measurement settings, or spatially varying, in which case we write $x_1(s),\ldots,x_p(s)$ for the observed covariates associated with the spatial process $Y(s)$ at location $s\in\mathcal S$. Covariates may affect both the marginal distributions and the spatial dependence structure. Their inclusion in the margins can be handled through  marginal or distributional regression models \citep[e.g.,][]{Hug2022} and is not considered  here \citep[see also][as a key reference for spatial regression models]{Cressie1993}. Likewise, global covariates can be incorporated into the dependence structure using  distributional-regression ideas. 
To the best of our knowledge, this has however found only limited attention in the literature. \citet{RieKleKne2023} propose a Gaussian-process regression model in which both the mean and the covariance structure depend on covariates, thereby inducing a Gaussian copula random field whose dependence structure varies with the global covariates.

In the presence of spatially varying covariates, a simple appraoch to incorporate them into the dependence structure is to let pairwise dependence depend not only on geographic separation, but also on differences in relevant covariates. For instance, if $x(s)$ denotes a spatially varying covariate such as elevation or temperature, the dependence between $Y(s_i)$ and $Y(s_j)$ may be determined jointly by $\|s_i-s_j\|$ and $|x(s_i)-x(s_j)|$. Equivalently, the original spatial domain may be embedded into a higher-dimensional covariate-augmented domain, for example through
\[
s
\longmapsto
\widetilde s
=
(s,\alpha x(s)),
\qquad
\alpha>0,
\]
and a copula random field may then be specified on the augmented domain
$
\widetilde{\mathcal S}
=
\{\widetilde s:s\in\mathcal S\}.
$
This is formally equivalent to defining a random field on a subset of $\mathbb R^q$ with $q>2$. Consequently, any copula-field construction in which pairwise dependence can be parameterized through a, not necessarily Euclidean, distance on the underlying domain remains valid after this embedding; see Section~\ref{subsec:top-down-copula-field} for examples and \citet{Schmidt2011,Reich2011} for analogous covariate-based constructions in classical spatial statistics. Such an augmented-distance construction is used explicitly in a copula context by \citet{ErhCzaSch2015}, with elevation differences serving as an additional measure of separation.

More generally, spatially varying covariates may enter pairwise dependence parameters directly, rather than only through an augmented distance. For example, if a copula-field model contains a parameter $\theta_{s_i,s_j}$ governing the dependence between $Y(s_i)$ and $Y(s_j)$, one may allow $\theta_{s_i,s_j}$ to depend on both the geographic distance $\|s_i-s_j\|$ and suitable symmetric summaries of the covariate values at the two locations. Thus, dependence may vary not only with spatial separation, but also, for instance, with the average elevation $\{x(s_i)+x(s_j)\}/2$. However, care is required, because an arbitrary pairwise specification need not be compatible with a valid copula random field, as discussed in the previous sections. Analogous constructions for covariate-driven nonstationary covariance models are developed by \citet{RisserCalder2015}.

\subsection{Further copula-based approaches in spatial statistics}\label{subsec:copula-y-space}
Copulas can be used not only to model the finite-dimensional dependence structure of a univariate spatial field but also to describe dependence among the components of a multivariate spatial field at a fixed location. To illustrate this second perspective, suppose that several quantities $Y_1(s),\ldots,Y_D(s)$ are observed at locations $s\in\mathcal S$, as in meteorological applications involving, for example, precipitation and temperature recorded at a network of stations. In this setting, $D$-variate copulas may be used to model the dependence among $Y_1(s),\ldots,Y_D(s)$ at a given location $s$, so that the copula itself may vary with $s$. More generally, let $x$ denote additional covariates, which may also vary across locations and, where appropriate, across response components. Copula-based distributional models \citep{KleKne2016, KocKle2025} specify the conditional joint cdf of $(Y_1(s), \dots, Y_D(s))$ given $x$ as
\[
F(y \mid x,s )=C(F_1(y_1|x,s),\ldots F_D(y_D|x,s)\mid x,s),
\]
where $y =(y_1, \dots, y_D)$.
Spatial variation may thus enter both the marginal distributions $F_j$ and the copula $C$, for example through anisotropic tensor-product splines \citep[see, e.g.,][for a recent contribution and further references]{BacKle2025} or Gaussian Markov random fields with possibly nonstationary covariance structures \citep{LinRueLin2011, LabGotKle2026}.

\section{Statistical inference for spatial models involving copulas} 
\label{sec:statistics-spatial-copulas}

In this section, we discuss key aspects of statistical inference for spatial models with copulas. The main interest here lies in estimation and prediction of the joint distribution built from a copula random field and suitable marginal models. 
As noted earlier, the marginal components often incorporate additional covariates, such as temporal trends, seasonal effects, or other spatially varying predictors, whereas the copula component captures the residual dependence structure after accounting for these marginal effects. Covariates may, however, also enter the copula field itself, as discussed in Section~\ref{subsec:copula-covariates}, thereby increasing the complexity of the resulting estimation problem.

\subsection{Estimation and inference}
The form of the available data is a key determinant of both the estimation strategy and the theoretical framework available to justify it. If repeated measurements are observed at a fixed finite set of locations, the data may be viewed as a sample of finite-dimensional spatial random vectors. Provided that the replicates are independent, inference can then be largely based on established methods from multivariate copula statistics; see Section~\ref{subsubsec:key-goals-methods-statistics-copulas} and, for a spatial factor-copula example, \citet{Krupskii2018}. If the replicates form a weakly dependent time series, corresponding results from time-series copula theory may be applicable, although the validity of particular procedures depends on the temporal dependence assumptions; see, for example, \citet{TanWanSunHer2019} and Section~\ref{sec:temporal-dependence}.

If, by contrast, only a single realization of $\mathbb Y$ is observed at a finite set of locations, dependence parameters cannot generally be estimated without additional structural assumptions. Common assumptions include stationarity, isotropy, or parametric dependence structures indexed by spatial distances or covariates. 
For instance, under stationarity information may be pooled across pairs of locations with similar spatial lags, while isotropy permits further pooling according to spatial distance alone; see also Section~\ref{subsec:estimating-copula-variograms}. 

Theoretical guarantees in the single-realization setting require genuinely spatial asymptotic arguments, typically under an increasing observation domain or an increasingly dense spatial design. Consistency and asymptotic normality have been obtained for some particular models and estimation procedures only, including composite likelihood methods for Gaussian-copula models with spatially clustered data \citep{Bai2014} and the Clayton-like copula random field \citep{Bevilacqua2024}. Nevertheless, a broadly applicable theory covering flexible spatial copula models, semiparametric marginal estimation, and corresponding inference remains limited. Hence, we summarize the main estimation approaches below at a deliberately general level.

Under either observation scheme, estimation may proceed jointly in a one-step approach or sequentially in a two-step approach; see Section~\ref{subsubsec:key-goals-methods-statistics-copulas} for further discussion. The specific method used for the copula component depends on the model class, the number of spatial locations, and the availability of a tractable likelihood. Whenever the joint copula density is available in explicit form and can be evaluated at reasonable computational cost, likelihood-based estimation is a natural choice. This includes many Gaussian copula models, factor copulas, and vine copulas of manageable dimension, although in the latter case sequential tree-by-tree estimation is often preferred in practice. Gaussian copula marginal regression provides a particularly relevant example, combining regression models for the margins with latent Gaussian dependence structures, including spatial correlation models \citep{Mas2012}. More generally, likelihood-based inference may become computationally demanding as the number of locations increases, or it may be unavailable when finite-dimensional densities do not admit a tractable form. Pairwise or, more generally, composite-likelihood methods then provide an important alternative; see \citet{Varin2011,Bel2015} for general overviews and \citet{Padoan2010,Bevilacqua2024} for specific spatial examples.

Bayesian methods provide an alternative framework for parameter estimation and uncertainty quantification.  For example,  \citet{San2010} propose a hierarchical Bayesian specification for Gaussian copula process models for spatial
extreme values.   When the full likelihood is unavailable, composite likelihoods may be used to construct approximate Bayesian procedures, although suitable adjustments are generally required to obtain appropriately calibrated posterior uncertainty; see, for example, \citet{Rib2012} for an application to spatial extremes. Recently, \citet{PeaGunCre2025} developed fully Bayesian copula-based spatial random-effects models for large, noisy, incomplete, and non-Gaussian spatial data.

Approximate Bayesian computation \citep[ABC;][]{SisFanBea2018} provides a likelihood-free alternative when an explicit likelihood is unavailable or computationally prohibitive to evaluate. Spatial applications commonly rely on summary statistics designed to capture relevant dependence features, such as empirical variograms, extremal coefficients, or tail-dependence summaries. In spatial extremes, for example, \citet{ErhSmi2012} showed that ABC based on suitably chosen summaries can provide useful inference for dependence parameters and may compare favorably with composite likelihood. More generally, \citet{Grazian2017} developed an ABC approach for estimating functionals of copula-based multivariate distributions.

\subsection{Spatial prediction}
Once fitted, copula models provide a natural framework for spatial prediction/interpolation \citep{Bardossy2008}. Indeed, let $Y_{0:d}=(Y(s_0), \dots, Y(s_d))^\top$, and suppose that we seek to predict $Y_0 = Y(s_0)$ based on a single observation of the subvector $Y_{1:d} = (Y(s_1), \dots, Y(s_d))^\top$. Assuming that $Y_{0:d}$ has a common marginal cdf $F$ (with density $f$) and copula $C_{0:d}$ (with density $c_{0:d}$), a straightforward calculation shows that the conditional distribution of $Y_0$ given $Y_{1:d}=y_{1:d}$ has the density
\[
f_{0 \mid 1:d}(y_0 \mid y_{1:d}) = f(y_0) \frac{c_{0:d}(F(y_0), \dots, F(y_d))}{c_{1:d}(F(y_1), \dots, F(y_d))},
\]
where $c_{1:d}(v_1, \dots, v_d) = \int_0^1 c_{0:d}(u,v_1, \dots, v_d)\, \diff u$ is the copula density of $Y_{1:d}$.
By a change of variables, the conditional mean $\mu(y_{1:d}) = \Exp[Y_0 \mid Y_{1:d} = y_{1:d}]$ can thus be written as
\[
\mu(y_{1:d}) = \int_0^1 F^{-1}(u) \frac{c_{0:d}(u, F(y_1),\dots, F(y_d))}{c_{1:d}(F(y_1), \dots, F(y_d))} \, \diff u,
\]
provided it exists.
This representation yields a practical interpolation scheme by replacing $F$ and $c_{0:d}$ with suitable estimators. Instead of the conditional mean, one may also use the conditional median; see \cite{Bardossy2008}. 

More generally, the conditional copula model from above provides the entire predictive distribution of $Y_0$ given the observations, allowing not only point predictions but also predictive intervals and other summaries of predictive uncertainty to be obtained. In practice, these predictive distributions are typically based on plug-in estimates of the marginal and copula parameters, while additional estimation uncertainty may be incorporated through bootstrap procedures or Bayesian posterior predictive inference. Theoretical guarantees for such inference, however, remain comparatively limited, particularly when inference is based on a single spatial realization.

These interpolation approaches are closely related to classical geostatistical ideas. In particular,  kriging \citep[Section~5.21]{Cressie1993} can be interpreted within a copula framework, a connection that is made explicit in \cite{KaziankaPilz2010}. More recent work has developed copula-based extensions of indicator kriging, such as copula-based multiple indicator kriging for non-Gaussian random fields \citep{Agarwal2021}, as well as related rank-based approaches to spatial prediction \citep{Juang2001}. Within the copula literature, additional flexibility can be achieved through combinations of copulas, for example via convex combinations of Archimedean copulas \citep{Sohrabian2021}.

Bayesian approaches were proposed by \cite{Kazianka2011}, while \cite{Alidoost2018} extend this framework to incorporate covariates.

\subsection{Implementations}
To conduct real data analyses, suitable software packages are essential. \cite{Wikle2019} provide a general introduction to spatio-temporal statistics using \textsf{R}. While considerable contributions with implementations  for Gaussian-process and latent-Gaussian models are available, comparatively few tools are suitable for flexible copula-based geostatistical modeling.  Among the established relevant packages, \texttt{SpatialExtremes} \citep{SpatialExtremes} provides extensive functionality for spatial extremes, including max-stable, copula-based, and hierarchical models.
The package \texttt{GeoModels} \citep{GeoModels} provides likelihood and composite-likelihood procedures for a range of Gaussian and non-Gaussian spatial and spatio-temporal random fields, including the Clayton-like copula field of \cite{Bevilacqua2024} discussed in Section~\ref{subsubsec:archimedean-copula-fields}.
More general copula modeling can be carried out using the package \texttt{copula} \citep{copula}, which implements a large collection of elliptical, Archimedean, and extreme-value copulas together with likelihood-based estimation, simulation, conditional distributions, and goodness-of-fit procedures. Although not specifically designed for geostatistical applications, the package \texttt{copula} can be combined with spatial covariance packages such as \texttt{geoR} \citep{RN-2001-013}, \texttt{RandomFields} \citep{randomfields}, or \texttt{gstat} \citep{gstat} to construct spatial copula models. For high-dimensional dependence structures, vine-copula software such as \texttt{VineCopula} \citep{vinecopula}  and \texttt{rvinecopulib} \citep{rvinecopulib} provides flexible pair-copula constructions, estimation, model selection, and conditional simulation. These packages can in principle be adapted to spatial settings as well. However, computational costs may be too high for large spatial dimensions. 
The \textsf{R} package \texttt{gcKrig} \citep{gcKrig} supports the analysis of geostatistical count data using Gaussian copulas.

To the best of our knowledge, currently available software offers only limited support for scalable estimation, prediction, and uncertainty quantification in spatial factor-copula models with replicated data.

\section{Spatio-temporal statistics using copulas}  \label{sec:temporal-dependence}

In many applications, the replicated-measurement setting described above naturally gives rise to spatio-tempo\-ral data, where observations are collected repeatedly over time at the same set of spatial locations \citep{Cressie2011}. The resulting observations may exhibit temporal dependence, leading to specific modeling and statistical challenges that will be discussed in this section. We note, however, that temporal dependence can sometimes be neglected in practice, for instance when observations consist of block maxima, so that temporal aggregation removes short-range dependence, or when successive measurements are sufficiently widely separated in time to be regarded as approximately independent.

The following notation will be used throughout: 
as before, the spatial domain is denoted by $\mathcal S \subset \R^2$ or $\mathcal S \subset \R^3$, while the temporal domain is written as $\mathcal T \subset \R$. In many applications, $\mathcal T$ is assumed to be discrete (for instance, for daily observations), and we may then assume without loss of generality that $\mathcal T \subset \N$. The target variable of interest at the space-time location $(s,t)$ is denoted by $Y_t(s)$, with respective cdf $F_{s,t}$.

As in Section~\ref{subsec:spatial-random-fields}, the stochastic properties of the space-time process $\mathbb Y^{(\mathrm{st})} = (Y_{t}(s))_{ (s,t) \in \mathcal S\times \mathcal{T}}$ are fully characterized by its \emph{space-time fidis} 
\[
F_{\mathcal I}(y) =  \Pr\big[ Y_t(s) \le y_{s,t} \, \forall (s,t) \in \mathcal I\big],
\]
where $y =(y_{s,t})_{(s,t) \in \mathcal I}$ and where $\mathcal I \subset \mathcal S \times \mathcal T$ ranges over all finite collections of space-time locations $(s,t)$. 
As in Section~\ref{sec:copula-random-fields}, this formulation naturally gives rise to \emph{space-time copula fidis} and corresponding \emph{copula processes} describing the full spatio-temporal dependence structure. Such approaches, and more generally approaches that employ copula models jointly across space and time, are discussed in Section~\ref{subsec:space-time-process-modeling}.

In practice, however, fully spatio-temporal copula models are often difficult to construct and estimate, particularly for flexible high-dimensional copula classes such as vine copulas. A common simplification is therefore to use copulas only for modeling the spatial dependence structure at fixed time points, while handling the temporal dependence separately. More specifically, writing 
$\mathbb Y_t^{\raisebox{-.2ex}{$\scriptstyle(\mathrm{s})$}}=(Y_{t}(s))_{s \in \mathcal S}$ 
for the spatial random field observed at time $t$, one obtains a time-indexed family of random fields to which the modeling approaches from the previous sections can be applied, potentially combined with suitable models for the temporal dynamics. Corresponding approaches are reviewed in Section~\ref{subsec:time-series-of-spatial-data}.

Finally, as in Section \ref{subsec:copula-covariates}, additional covariates may be observed, and spatial location or time may itself be treated as a covariate. Since the approaches discussed in Section~\ref{subsec:copula-covariates} directly carry over to the space-time case we do not repeat them here.

\subsection{Joint spatio-temporal modeling using copulas}
\label{subsec:space-time-process-modeling}

A \emph{spatio-temporal copula process} is a stochastic process $\mathbb U^{(\mathrm{st})}=(U_t(s))_{t \in \mathcal T,\, s \in \mathcal S}$ whose univariate marginal distributions are standard uniform. For any such process $\mathbb U^{(\mathrm{st})}$ and any prescribed collection of marginal cdfs $F_{s,t}$ (possibly depending on additional covariates), the process
\[
\mathbb Y^{(\mathrm{st})} = (Y_t(s))_{(s,t) \in \mathcal S\times \mathcal{T}},
\qquad
Y_t(s)=F_{s,t}^{-1}(U_t(s)),
\]
inherits the copula fidis of $\mathbb U^{(\mathrm{st})}$ and has marginal cdfs $F_{s,t}$. As in the purely spatial setting, this construction separates the marginal modeling from the spatio-temporal dependence modeling and thereby allows for considerable modeling flexibility.

\subsubsection{Implicit modeling}
As in Section~\ref{subsec:top-down-copula-field}, spatio-temporal copula processes can be constructed implicitly from established random fields through marginal standardization. More precisely, any spatio-temporal process $\mathbb Z^{(\mathrm{st})} = (Z_t(s))_{t \in \mathcal T,\, s \in \mathcal S}$ with continuous marginal cdfs $G_{s,t}$ induces a spatio-temporal copula process $\mathbb U^{(\mathrm{st})}$ through $U_t(s) = G_{s,t}(Z_t(s))$, whose copula fidis are Kolmogorov consistent. Common starting points for $\mathbb Z^{(\mathrm{st})}$ are spatio-temporal Gaussian or $t$-processes, which are typically specified through their covariance functions. This formulation facilitates the incorporation of structural dependence assumptions by imposing, for instance, spatial or temporal stationarity, spatial isotropy, or separability on the covariance function; see, for example, \cite[Section~6.1]{Cressie2011}. We also refer to \cite{Gne2002} for a collection of widely used parametric covariance models. 
Analogous implicit constructions are common in the modeling of spatio-temporal extremes, where max-stable and Pareto processes serve as the primary latent process-level models; see, for instance, \citet{DavisonHuser2014,HusWad2020,DellOro2025} for recent reviews and developments.
Finally, non-stationary space--time models based on the SPDE approach of \citet{LinRueLin2011} were proposed by \citet{CLAROTTO2024100847}.

\subsubsection{Modeling based on non-monotone mappings}
Approaches such as those discussed in Section~\ref{subsubsection:new-from-old-via-non-monotone-mappings} are, at least conceptually, also applicable in the spatio-tem\-po\-ral setting. In particular, non-monotone marginal transformations of spatio-temporal Gaussian or t-pro\-cesses again yield Kolmogorov consistent spatio-temporal copula processes that admit parameterizations in terms of covariance functions. To the best of our knowledge, however, such constructions, as well as their connections to specific structural assumptions in space-time statistics, have not yet been studied systematically in the literature.

\subsubsection{Space-time factor-copula models}
\label{subsubsec:space-time-factor-copulas}
Factor copula random fields described in Section~\ref{subsubsec:factor-copula-fields} have been extended to the spatio-temporal setting in \cite{KruGen2017}. The construction is based on a latent space-time Gaussian field $Z_t(s)$ and a latent common factor process $V_t(s) = \alpha(s,t)\mathcal{E}_{\mathcal{P}(t)}$ that affects observations across both space and time, thereby inducing flexible spatio-temporal dependence structures that allow for tail dependence and asymmetry. The process $\mathcal{P}(t)$ is a Poisson process with time-varying intensity function. As in the spatial case, the factor copula random field $U_t(s)$ is obtained by marginal standardization of the process $W_t(s) = Z_t(s) + V_t(s)$. 
Spatial isotropy and temporal stationarity arise as special cases when $\alpha$ and the intensity function are invariant with respect to space and time.
Another appealing property of the model is that it remains computationally tractable, since the likelihood can often be evaluated in closed form or via low-dimensional integration. This enables efficient maximum-likelihood estimation and simulation.

\subsubsection{Copula-based space-time Markov models}
Structural assumptions in the form of Markovian properties provide a fundamentally different approach to fully copula-based dependence modeling. This line of research originates from the foundational work of \cite{Darsow1992} and has subsequently been applied successfully in time series analysis; see, for instance, \cite{Chen2006}. An extension that additionally accounts for space was proposed by \cite{TanWanSunHer2019} for finite spatial domains, although the underlying construction extends more generally.
In the simplest setting, such an extension assumes that the copula random field $(U_t(s))_{s \in \mathcal S}$ is stationary and first-order Markovian as a function of $t \in \mathcal T=\N$. In that case, the stochastic properties of the spatio-temporal copula process $\mathbb U^{(\mathrm{st})}$ are fully determined by the joint law of the consecutive fields $ (U_1(s))_{s \in \mathcal S},(U_2(s))_{s \in \mathcal S})$, which in turn is characterized by the associated copula fidis
\[
C_{\mathcal I}(u) = \Pr\bigl[ U_t(s) \le u_{s,t} \ \forall (s,t) \in \mathcal I\bigr], 
\]
where $u =(u_{s,t})_{(s,t) \in \mathcal I}$ and where $\mathcal I = \mathcal S_d \times \{1,2\}$ for some finite subset $\mathcal S_d \subset \mathcal S$ of cardinality $d$. 
The specific constructions for $C_{\mathcal I}$ proposed by \cite{TanWanSunHer2019} are primarily based on non-monotone marginal transformations of Gaussian copulas in the spirit of \cite{Bardossy2006} (see Section~\ref{subsubsection:new-from-old-via-non-monotone-mappings}), combined with suitable parametric covariance models.

\subsubsection{Space-time vine copulas}

\citet{Grler2012} proposed an extension of the spatial C-vine construction from Section~\ref{subsec:bottom-up} to the spatio-temporal setting. Specifically, for a fixed maximal temporal lag $L\in\N$, they assume that the spatio-temporal process is temporally stationary and model the dependence between the current observation vector
$
Y_{0:d}^{t}
=
(Y_t(s_0),\dots,Y_t(s_d))^\top
$
and the lagged vectors
$
Y_{1:d}^{t-1}, \dots, Y_{1:d}^{t-L}
$
through an $(Ld+d+1)$-dimensional C-vine copula rooted at $Y_t(s_0)$. The construction thereby captures both  spatial dependence and temporal dependence up to lag $L$ within a unified copula framework. The bivariate copulas linking $Y_t(s_0)$ and $Y_{t-\ell}(s_i)$, for $i\in\{1,\dots,d\}$ and $\ell\in\{0,\dots,L\}$, are parameterized as functions of the spatial distance $\|s_i-s_0\|_2$ and the temporal lag $\ell$. The temporal stationarity assumption is achieved through  copula parameters that depend only on the lag $\ell$, rather than on the absolute time index $t$. This parameterization yields a parsimonious model and thereby makes prediction at unobserved spatial locations and future time points feasible through the methodology outlined in the previous sections.

Extending earlier work by \cite{BeareSeo2015} and \cite{Smith2015}, \cite{Nagler2022} consider more general R-vine copula constructions for jointly modeling spatial and temporal dependence. Specifically, they derive conditions on the underlying tree structure that ensure temporal stationarity of the resulting model, develop methods for estimation, simulation, and prediction, and establish corresponding asymptotic theory. Their framework does not explicitly incorporate spatial information such as distances or covariates, an aspect whose integration appears to be a promising direction for future research.

\subsection{Copulas for spatial dependence in time series}
\label{subsec:time-series-of-spatial-data}

In time-series analysis, one common strategy to capture spatial dependence  is  through a two-step approach: first, temporal models are fitted separately for each marginal series (i.e., at each location), after which copula-based methods are applied to capture the spatial dependence structure of the resulting residuals. The second step can then, in principle, be carried out using any of the spatial models and methodologies described in the previous sections. 

An instructive example is provided by \citet{ErhCzaSch2015}, who consider daily mean temperature measurements collected at 54 locations over a period of three years. To account for temporal dependence, the authors specify for each marginal time series an autoregressive model of the form $Y_t(s)=\mu_t(s)+\eps_t(s)$, where the mean function is given by $\mu_t(s)=g(s,t,Y_{t-1}(s),Y_{t-2}(s))$ for some parametric function $g=g_\theta$ that additionally captures annual seasonality effects. The error vectors $\bm\eps_t=(\eps_t(s_1),\dots,\eps_t(s_d))^\top$ are assumed to be iid across time, with skew-$t$ margins and a truncated R-vine copula dependence structure as described in Section~\ref{subsec:bottom-up}. After fitting the marginal models, the copula parameters are estimated from the residual vectors $\hat{\bm\eps}_t = (\hat\eps_t(s_1), \dots, \hat\eps_t(s_d))^\top$, where $\hat\eps_t(s)=Y_t(s)-\hat g(s,t,Y_{t-1}(s),Y_{t-2}(s))$. A similar strategy, albeit based on different marginal models, was employed by \cite{Pereira2016, Pereira2017} for streamflow simulation in hydroelectric power applications.  \citet{Shahriari2025} apply a Bayesian autoregressive time series model with temporal and spatial correlations in transport planning.

\section{Summary and outlook}
\label{sec:discussion-outlook}

Copula-based approaches offer both a useful conceptual perspective on spatial dependence and a flexible framework for developing models beyond the classical Gaussian paradigm. Viewing existing spatial methods through a copula lens can clarify how their dependence structures operate, while genuinely copula-based constructions expand the range of models available for spatial data. Extending copula methodology from finite-dimensional vectors to random fields, however, requires additional structural constraints, including compatibility across locations and the ability to represent key spatial features such as stationarity, isotropy, anisotropy, and nonstationarity.  Despite a growing body of work, the current repertoire of copula random fields remains comparatively small, especially when contrasted with the rich literature on finite-dimensional copula models. This gap suggests substantial opportunities for future methodological developments, several of which we discuss below.

\smallskip
\emph{Flexibility and scalability.}
Most spatial copula-field models are built from established random-field constructions, often with Gaussian random fields as their starting point. By contrast, for many asymmetric and non-elliptical finite-dimensional copula models, including Archimedean and vine copulas, it remains unclear how to construct flexible and valid spatial extensions over continuous or infinite domains. Important challenges for such extensions, and for spatial copula modelling more generally, include interpretability and scalability, with the latter referring to computational feasibility in terms of runtime and memory as the number of observations, spatial locations, or model parameters increases.

\smallskip
\emph{Asymptotic theory and validated inference.}
Although established multivariate or time-series copula theory can often be applied when spatial observations are available as independent or weakly dependent replicates at a fixed set of locations, broadly applicable asymptotic theory remains limited for genuinely spatial observation schemes; see Section~\ref{sec:statistics-spatial-copulas}. This is particularly true when inference is based on a single realization observed over an increasing domain or at an increasingly dense collection of locations, while further challenges arise from high-dimensional dependence models, composite likelihoods, estimated margins, and temporally dependent replicates. Developing more general results on consistency, limit distributions, inference, and the validity of resampling methods therefore remains an important direction for future research. Similar open research questions also arise when resorting to Bayesian inference.

\smallskip
\emph{Missing and misaligned data.}
Missing and spatially misaligned observations are common in geostatistical applications and can substantially affect estimation and uncertainty quantification \citep{GelDigGut2010,FinBanCoo2014}. Missingness arises when response or covariate values are unavailable, whereas misalignment occurs when variables are observed at different spatial locations; although closely related, the two settings may involve different stochastic mechanisms. Standard approaches often rely on missing-at-random assumptions, but these may be inappropriate when observation availability or location is related to the underlying process, as may occur because of inaccessible terrain, ecologically sensitive areas, or process-dependent measurement limitations. Fully probabilistic models, in particular Bayesian ones,  provide a natural framework to jointly account for missingness, misalignment, and the associated uncertainty \citep{PeaGunCre2025}, although their inferential consequences remain insufficiently understood. These challenges are not specific to copula models, but copulas can themselves support imputation; for example, \citet{CHAPON2023100591} use a vine copula to reconstruct a target-station time series from neighboring stations while accommodating missing observations throughout the network.

\smallskip
\emph{Deep learning.} 
A potentially promising direction for future research lies at the intersection of copula modeling, spatial statistics, and deep learning. Neural methods for spatial inference, prediction, and simulation have gained increasing interest recently. Examples include neural Bayes estimators and graph-neural-network-based inference for spatial process models \citep{SaiZamHus2024, SaiZamRicHus2025}, neural conditional simulation for complex spatial processes \citep{WalZamHusKuu2025}, neural operator approaches for spatio-temporal forecasting \citep{NagZamSinCre2026}, and normalizing-flow-based models for nonstationary spatial processes \citep{NagZamSun2025}. While these methods are typically developed independently of copula theory, they demonstrate that neural networks can successfully learn high-dimensional  dependence structures and provide scalable alternatives to traditional likelihood-based inference. 

Conversely, several generative approaches explicitly combine copulas with neural networks. Vine copula autoencoders model the lower-dimensional latent representation learned by an autoencoder using a vine copula, thereby yielding a flexible high-dimensional generative model \citep{TagAckVat2019}. Copula flows instead use normalizing flows to learn both the marginal distributions and the copula density \citep{KamAssDei2021}. Copula-based generative models may also be combined with neural methods less directly, for example by generating synthetic weather and climate data to augment the training of neural-network emulators \citep{MeyNagHog2021}. Further deep-learning developments in the copula literature include neural inference procedures for copula state-space and time-series models \citep{FynGunZam2026}; see \citet{CobGroLiuWen2026} for an extensive recent review and further references. Extending such approaches to spatially indexed data may provide a fruitful route toward flexible and scalable copula-based spatial models.

%
%

\section*{Declarations}
During the preparation of this manuscript, the authors used ChatGPT, developed by OpenAI, to improve the language and clarity of the text. The authors reviewed and edited all AI-assisted content and take full responsibility for the final manuscript.

\begin{funding}
Both authors were supported by the Deutsche For\-schungsgemeinschaft (DFG, German Research Foundation; Project-ID 520388526;  TRR 391:  Spatio-temporal Statistics for the Transition of Energy and Transport). Nadja Klein also acknowledges funding through the DFG Emmy Noether grant KL3037/1-1.
\end{funding}

\bibliographystyle{imsart-nameyear} 
\bibliography{biblio.bib}       


\end{document}